# Tidal Currents Detected in Kraken Mare Straits from *Cassini* VIMS Sun Glitter Observations

MICHAEL F. HESLAR,[1] JASON W. BARNES,[1] JASON M. SODERBLOM,[2] BENOIT SEIGNOVERT,[3] RAJANI D. DHINGRA,[3] AND CHRISTOPHE SOTIN[3]

[1]*Department of Physics, University of Idaho*
*Moscow, ID 83843-0903*
[2]*Department of Earth, Atmospheric, and Planetary Sciences, Massachusetts Institute of Technology*
*Cambridge, MA 02139*
[3]*Jet Propulsion Laboratory, California Institute of Technology*
*Pasadena, CA 91109*

## Abstract

We present *Cassini* VIMS observations of sun glitter – wave-induced reflections from a liquid surface offset from a specular point – on Kraken Mare. Sun glitter reveals rough sea surfaces around Kraken Mare, namely the coasts and narrow straits. The sun glitter observations indicate wave activity driven by the winds and tidal currents in Kraken Mare during northern summer. T104 *Cassini* VIMS observations show three sun glitter features in Bayta Fretum indicative of variegated wave fields. We cannot uniquely determine one source for the coastal Bayta waves, but we lean toward the interpretation of surface winds, because tidal currents should be too weak to generate capillary-gravity waves in Bayta Fretum. T105 and T110 observations reveal wave fields in the straits of Seldon Fretum, Lulworth Sinus, and Tunu Sinus that likely originate from the constriction of tidal currents. Coastlines of Bermoothes and Hufaidh Insulae adjoin rough sea surfaces, suggesting a complex interplay of wind-roughened seas and localized tidal currents. Bermoothes and Hufaidh Insulae may share characteristics of either the Torres Strait off Australia or the Aland region of Finland, summarized as an island-dense strait with shallow bathymetry that hosts complex surface circulation patterns. Hufaidh Insulae could host seafloor bedforms formed by tidal currents with an abundant sediment supply, similar to the Torres Strait. The coastlines of Hufaidh and Bermoothes Insulae likely host ria or flooded coastal inlets, suggesting the Insulae may be local peaks of primordial crust isolated by an episode of sea-level rise or tectonic uplift.

*Keywords:* sun glitter, Kraken Mare, tidal currents, exo-oceanography, coastal morphology

## 1. INTRODUCTION

The *Cassini* mission peered through Titan's hazy veil, discovering isolated liquid bodies on its surface (Stofan et al. 2007). The majority of surface liquids reside in the hydrocarbon seas (maria) of the north polar latitudes (Mitri et al. 2007, Stofan et al. 2007). The hydrological cycle of liquid methane is particularly active in the presence of these vast hydrocarbon seas, including the seasonal evolution of wind and precipitation patterns (Tokano et al. 2009). In addition, Titan's orbital eccentricity imposes tidal effects on the atmospheric winds (Tokano & Neubauer 2002) and sea circulation (Dermott & Sagan 1995). We can observe the meteorological conditions and tidal effects as surface waves on the largest hydrocarbon seas.

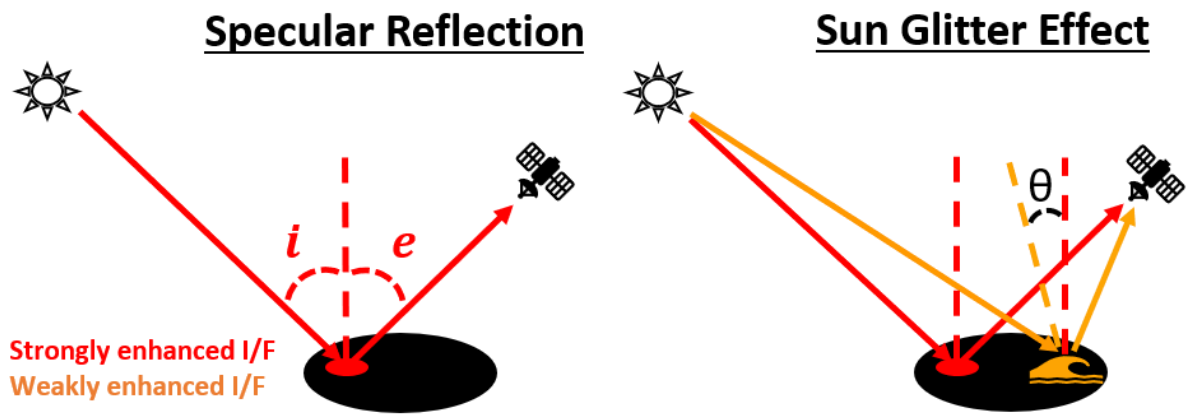

Figure 1. This cartoon illustrates the observation geometry of a specular reflection (sunlight directly reflected off a planetary surface) and sun glitter (waves tilting the liquid surface toward an observer, like *Cassini* VIMS) off a dark Titan sea surface in the left and right diagrams respectively. The dashed lines indicate the local surface normals for flat and rough liquid surfaces (red and orange respectively). The incidence and emission angles are *i* and *e*, which are in the same plane and equal to each other for both cases. Sun glitter retains a specular geometry but instead tilts the surface normal by $\theta$, known as the specular deviation angle. Red and orange colors represent strongly-enhanced radiance from a flat sea surface compared to weakly-enhanced radiance from a rough sea surface. I/F is the ratio of observed to incident flux measured by the *Cassini* Visual and Infrared Mapping Spectrometer (VIMS).

On Earth, oceanographers often observe surface waves over the seas and oceans from remote sensing (Breon & Henriot 2006). Surface waves represent macroscopic sea surface roughness that cause a broad, specular surface reflectance phase function known as sun glitter. Sun glitter imagery can reveal important details about the local environmental conditions on the surface, such as wind speeds (Breon & Henriot 2006) and ocean currents (Kudryavtsev et al. 2017b). Oceanographers can also infer information about oceanographic processes beneath the sea, such as the stratification of water masses (Jackson 2007). Similar sun glitter observations could yield useful insights in the oceanographic processes on Titan, which may rival the complexity in Earth's oceans (Lorenz 2013).

One major consequence from the summertime meteorological activity over the lakes and seas is the potential development of waves. Titan wind-wave models suggested that winds would reach sufficient speeds for capillary wave action during the late spring (Hayes et al. 2013, Lorenz & Hayes 2012).

Liquid bodies act like mirrors when liquids have sea surface heights smaller than the observing wavelength of the spectrometer (Barnes et al. 2011a). Satellites see blinding specular reflections of the Sun when in the appropriate observational geometry. As Figure 1 illustrates, a satellite observes sun glitter as bright spots with enhanced reflectivity in the vicinity of the specular point.

Sun glitter often occurs in isolated regions, separate from the specular reflection, where the sea surface is rougher (i.e. larger sea surface height anomalies) and some facets tilt toward the observer (*Cassini*). We quantify the sea surface roughness by the specular deviation angle ($\theta$ in Figure 1), which describes the angle at which the liquid surface needs to be tilted toward an observer to achieve a specular reflection.

For Titan, the Visual and Infrared Mapping Spectrometer (VIMS) (Brown et al. 2004) observes solar specular reflections as bright or possibly saturated pixels in the atmospheric windows used for the color image of Figure 2. In particular, the 5 $\mu$m window shows a preferential enhancement in brightness (Sotin et al. 2012, Stephan et al. 2010), indicating the reflector is at the surface. The specular reflection is accompanied by an aureole in the surrounding annulus of pixels (Barnes et al. 2013, Barnes et al. 2014, Stephan et al. 2010). We denote this region of intense reflectance and forward-scattering (Tomasko et al. 2008) with specular deviation angles less than two degrees as the "specular zone" in this paper. We also define a "sun glitter zone" as an annulus around the specular zone where sun glitter can be easily detected as regions of weakly enhanced reflectance relative to the darker liquid surface, shown in Figure 2. The sun glitter zone is likely confined to a specular deviation angle of ~9-10° due to the similar steepness ratio (wave height to wavelength) for breaking waves in liquid methane and water (Craik 2004, Hayes et al. 2013).

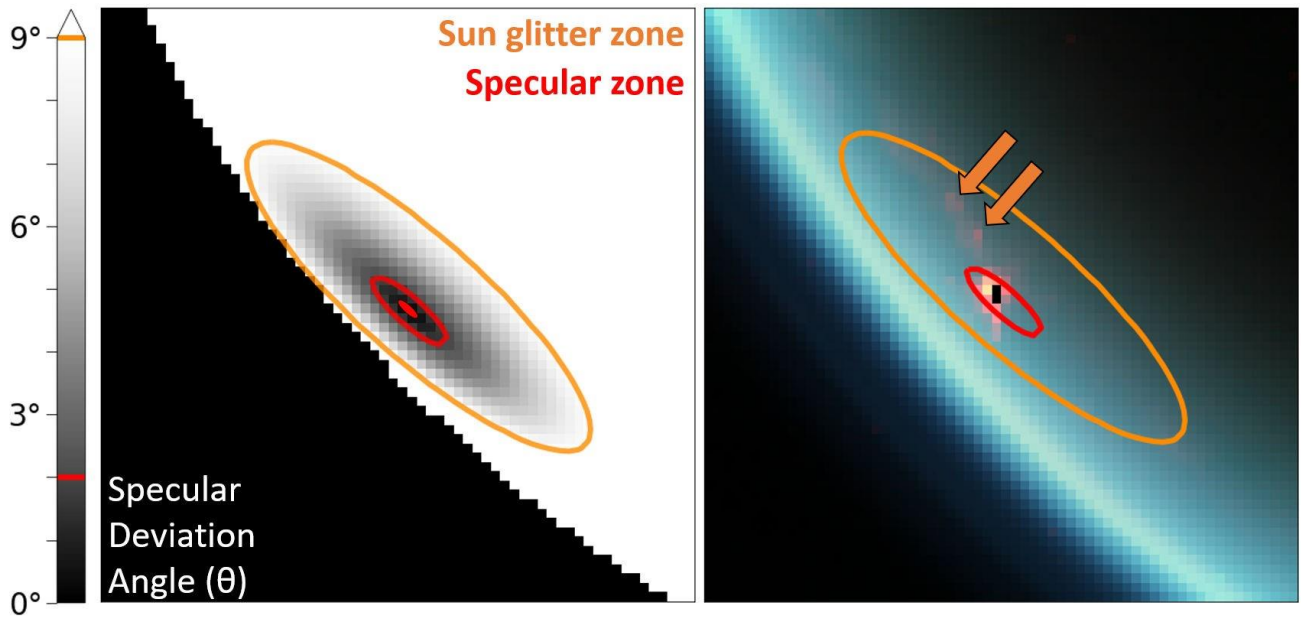

Figure 2. The left image shows a modeled specular deviation angle ($\theta$) map for the first-documented VIMS cube with sun glitter, CM_1721848119_1, modified from Barnes et al. (2014). The right image shows the corresponding cube with a wet-sidewalk color scheme (RGB: 5.0, 2.03, 2.79 $\mu$m) (Dhingra et al. 2019). The red and orange ellipses represent the theoretical specular and sun glitter zones ($\theta = 2$ and 9° respectively). The orange arrows identify the brightened pixels appearing in the sun glitter zone, where VIMS is most sensitive to sun glitter. In the left image, a filled red ellipse indicates the theoretical specular point, which is observed as saturation in the two black pixels of the right image.

Initial surveys of specular reflections on Ontario Lacus, Jingpo Lacus, and Ligeia Mare turned out to be very smooth and lacked any signature of waves (Barnes et al. 2011a, Wye et al. 2009, Zebker et al. 2014). A T59 specular reflection observed over Kraken Mare produced an inconclusive light curve, but wind or wave activity was postulated (Barnes et al. 2011a). In the late northern spring, Barnes et al. (2014) found several isolated bright pixels on Punga Mare during the T85 flyby (northern spring), providing the first definitive evidence of extraterrestrial waves. Soon after, the detection of RADAR-bright "magic islands" in Ligeia Mare were also theorized to be surface waves (Hofgartner et al. 2014, Hofgartner et al. 2016). Recent findings suggest Kraken Mare sea surface roughness constraints at 6-10 mm from RADAR altimeter measurements (Grima et al. 2017) and <3.6 cm from isolated bistatic RADAR observations (Marouf et al. 2016).

This paper investigates the occurrence of waves in constricted straits of the maria to probe oceanic processes, including tides and wind activity. Section 2 describes moderate-sampling (∼17-21 km) sun glitter observations in large areas of Kraken Mare as observed by specific near-infrared (NIR) atmospheric windows of Titan's atmosphere during the T104, T105, and T110 Titan flybys. Section 3 describes our examination of VIMS IR spectra that provide evidence to distinguish sun glitter from other surface and atmospheric features. Section 4 includes an analysis of fine-sampling sun glitter in select straits (Freta) of Kraken Mare. Section 5 considers the possible scenarios for consistent wave action and their implications on oceanographic phenomena in Kraken Mare, before we conclude in Section 6.

## 2. MODERATE-SAMPLING SUN GLITTER OBSERVATIONS

We focus our search for sun glitter in the later Titan flybys (after T100) that cover the northern spring and summer seasons with anticipated increases in capillary-gravity wave activity (Hayes et al. 2013, Lorenz & Hayes 2012). The northern spring and summer also offered the best chances to observe sun glitter with the constant daytime over the maria. We include some VIMS observation at coarse sampling (≥10 km/pixel) with bright specular reflections from the T104 and T105 flybys in Table 1. The T104 and T105 flybys were accompanied with fine-sampling sun glitter observations (<10 km/pixel) over the same areas of Bayta and Seldon Freta.

In Figure 3, we showcase a color mosaic of several VIMS observations of the largest sea of Titan, Kraken Mare, collected from the T104 flyby on 2014 August 21. We display the data using a wet-sidewalk RGB color scheme with the three most transparent atmospheric windows of Titan: 5.0, 2.03, and 2.79 $\mu$m (Dhingra et al. 2019). The left binary map in Figure 3 shows the relevant features that we discuss in this paper. Close flyby distances (≤50,000 km) to Titan allow us to observe details (≤25 km/pixel) of the surface and tropospheric features in the Titan sea district. The notable bright feature of the T104 mosaic in Figure 3 is a specular reflection and associated aureole caused by forward-scatter from haze (Barnes et al. 2013), which is located north of Hufaidh Insulae. The regions that appear bright pink in this color scheme (orange arrows in Figure 3) are areas of sun glitter because they are located away from the specular point and on known liquid surfaces.

We overlaid polar-projected VIMS observations averaged over the 5 $\mu$m window onto a global ISS map of Titan (Karkoschka et al. 2017) to put the specular reflection and sun glitter into a geographical context in Figure 4. The 5 $\mu$m window has the highest transmission (>90%), which allows for the most detailed surface studies of Titan (Sotin et al. 2012). We use Delaunay triangulation to interpolate the VIMS observation into a polar projection and retain the details of the raw data (Le Mouelic et al. 2019). We isolate the sun glitter zone with specular deviation angles between 2 and 9 degrees. Observational conditions of sun glitter, including spacecraft altitude (Soderblom et al. 2012), airmass (Le Mouelic et al. 2019), and the specular deviation angle (Cox & Munk 1954), vary for each Titan flyby and are corrected for an accurate comparison in Figure 4. We remove false positives (i.e. cosmic ray pixels) with a 3×3 low-pass filter if the pixels are only bright in a few neighboring spectral channels.

In the moderate-sampling 5 $\mu$m observations of Figure 4, the specular zone is confined to a small region of bright pixels that contains the specular point. In the sun glitter zone, we find several instances of sun glitter up to 350 km away during both T104 and T105 flybys. In particular, the sun glitter overlays the same locations in Kraken Mare, Bayta and Seldon Freta (narrow straits). Note that the large red feature in the T105 observation in Figure 4 is likely sun glitter, located to the east of Penglai Insula. The detection of sun glitter on two separate flybys may be indicative of consistently rough sea surfaces in the Freta of Kraken Mare. The presence of sun glitter is not surprising for coastal areas that are favorable for wave generation from turbulent surface currents (Job 2005, Lorenz 2014), yet prior NIR and RADAR specular observations indicated smooth sea surfaces (Barnes et al. 2011a, Hayes 2016).

Table 1. Details on select VIMS observations with sun glitter for the T104, T105, T106, and T110 flybys. Each cube name starts with "CM_". SC Alt. is shorthand for spacecraft altitude. We denote the relative quality of the VIMS observation as "Fine" if its mean spatial sampling is less than 10 km/pixel. $f$ represents true anomaly in degrees.

| Flyby | Cube CM_ | Image mid-time Date, Time [UTC] | Sampling [km/pixel] | Phase [°] | Incidence [°] | Emission [°] | SC Alt. [km] | $f$ [°] | Sampling Quality |
|-------|----------|---------------------------------|---------------------|-----------|---------------|--------------|--------------|---------|------------------|
| T104 | 1787311929_1 | 2014-08-21, 10:50:46 | 24 | 103 | 24-85 | 17-89 | 44,667 | 249 | Coarse |
| T104 | 1787310233_1 | 2014-08-21, 10:04:53 | 17 | 102 | 43-58 | 44-60 | 35,502 | 248 | Coarse |
| T104 | 1787307197_1 | 2014-08-21, 09:20:28 | 10 | 99 | 58-77 | 24-51 | 19,114 | 247 | Coarse |
| T104 | 1787306033_1 | 2014-08-21, 08:54:53 | 7 | 97 | 57-62 | 35-40 | 12,879 | 247 | Fine |
| T105 | 1790066425_1 | 2014-09-22, 07:45:01 | 21 | 109-111 | 23-77 | 37-89 | 42,680 | 248 | Coarse |
| T105 | 1790059135_1 | 2014-09-22, 05:39:57 | 2 | 86-90 | 50-55 | 31-41 | 4,055 | 246 | Fine |
| T105 | 1790059235_1 | 2014-09-22, 05:43:45 | 3 | 89-97 | 48-56 | 34-49 | 4,506 | 246 | Fine |
| T106 | 1792827131_1 | 2014-10-24, 06:38:12 | 36 | 120 | 34-79 | 40-88 | 77,873 | 250 | Coarse |
| T110 | 1805210863_1 | 2015-03-16, 14:30:21 | 2 | 75-108 | 49-53 | 58-66 | 2,408 | 246 | Fine |
| T110 | 1805212073_1 | 2015-03-16, 14:49:48 | 3 | 133-135 | 58-73 | 62-75 | 8,219 | 246 | Fine |

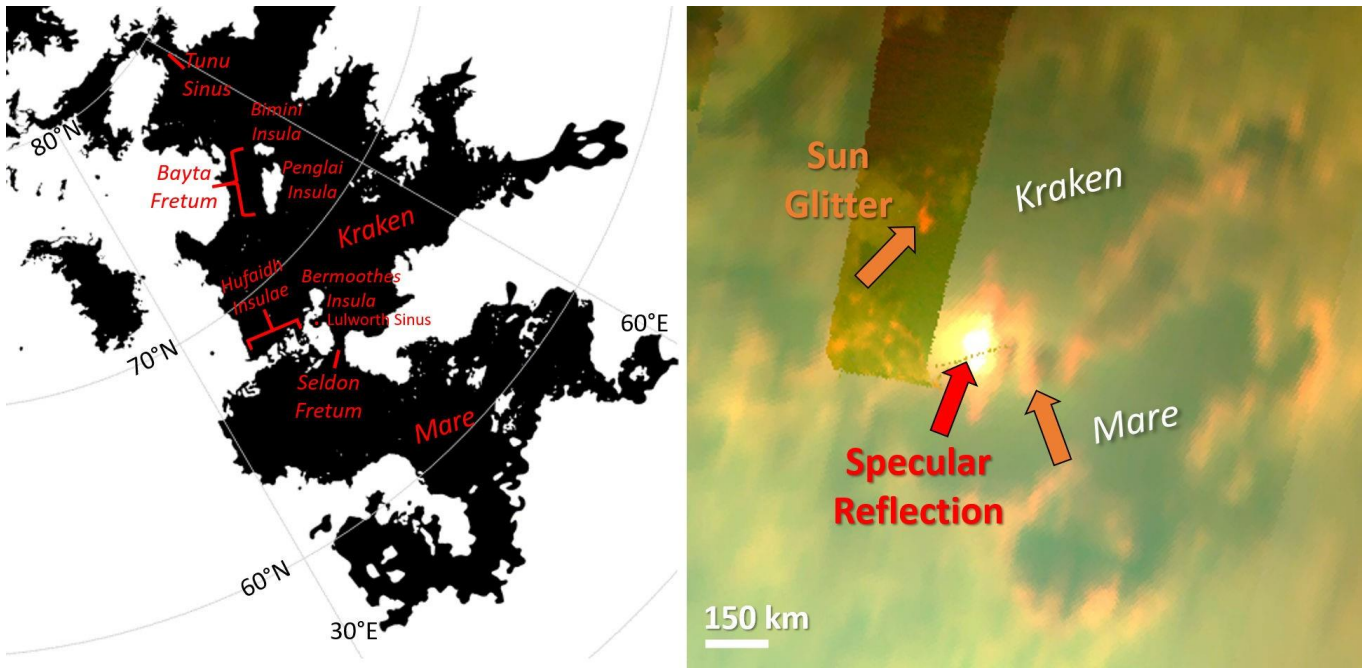

Figure 3. The left image shows a binary map of the Titan maria, derived from *Cassini* Imaging Science Subsystem (ISS) and RADAR data, and named features relevant for this publication. The right image shows the *Cassini* VIMS north polar map from the T104 flyby cropped onto Kraken Mare in the wet-sidewalk color scheme of Figure 2. Liquid surface features, specifically the specular reflection and sun glitter (red and orange arrows), appear as anomalously bright, uniform spots in the narrow straits or Freta of Kraken Mare.

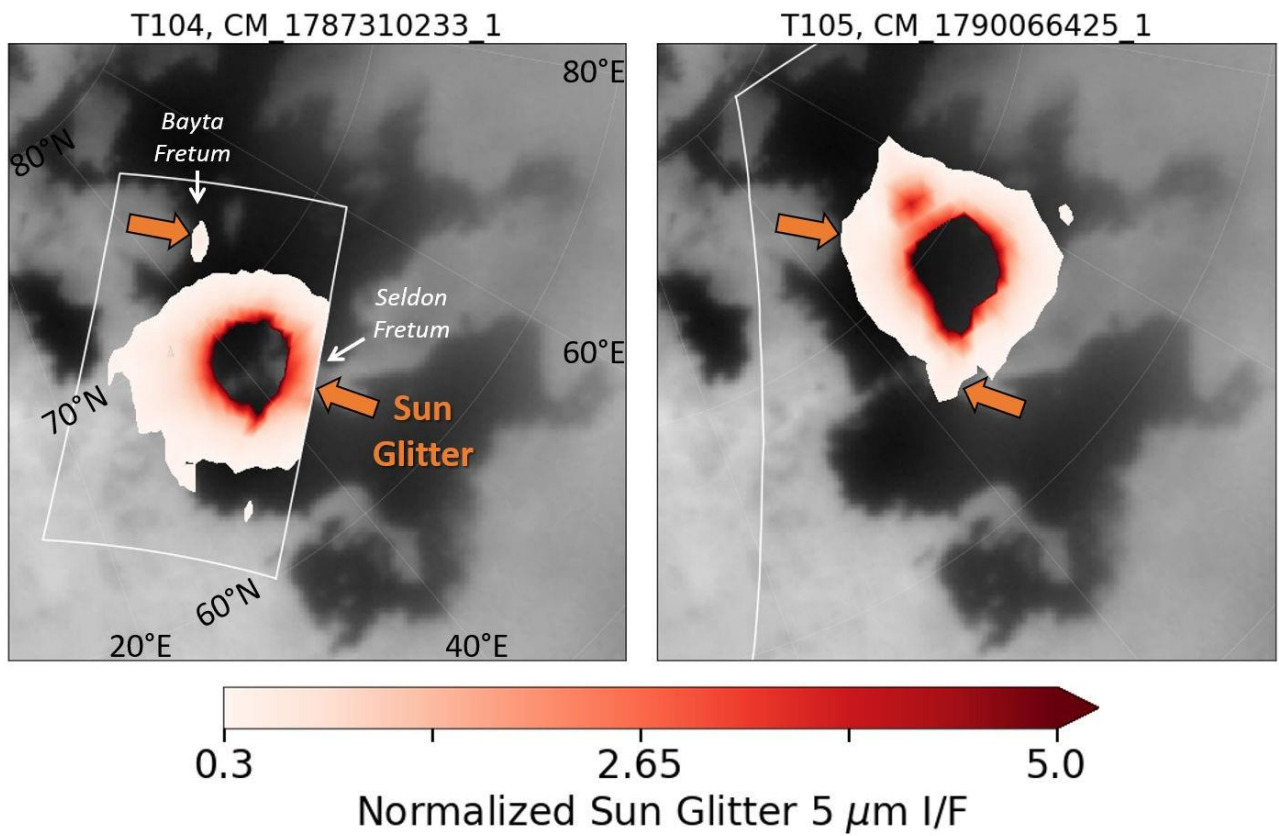

Figure 4. Probable regions of sun glitter are shown in the coarse-sampling VIMS cubes of Kraken Mare (~17 and 21 km/pixel respectively) for the T104 and T105 flybys, CM_1787310233_1 and CM_1790066425 1, in a normalized 5 $\mu$m polar projection. The 5 $\mu$m I/F is normalized by three factors: 1) the airmass used in Le Mouelic et al.(2019), 2) a quadratic dependence of the specular deviation angle at the edge of the sun glitter zone ($\theta = 9°$), and 3) a quadratic dependence of the spacecraft altitude at 44,000 km. The cubes are overlaid on an ISS basemap of Kraken Mare (Karkoschka et al. 2017). The translucent grid lines represent lines of latitude and longitude. Sun glitter overlays Bayta and Seldon Freta (orange arrows) during both flybys, indicating persistent roughness in these regions.

We note that the morphology of the observed sun glitter zone can provide insights to the overall sea state of Kraken Mare. Mainly, the T105 zone shows a larger areal extent and more elliptical profile. The Seldon sun glitter is brighter during the T104 flyby, while Bayta sun glitter shows similar I/F on both flybys. Overall, the sun glitter observations between the T104 and T105 flybys are suggestive of a variable, rough sea state in the north Kraken Mare region on two summer Titan days.

## 3. SPECTRAL ANALYSIS

We derive the spectral information of various features using individual VIMS spectra in this section. In Figure 5, we compare the spectra of the notable surface features, namely the hydrocarbon sea, evaporite deposits, sun glitter, and a specular reflection, from T104 VIMS cube CM_1787311929_1 due to its large spatial coverage at moderate sampling (24 km/pixel). The main atmospheric feature, tropospheric clouds, was observed in the T104 VIMS cube CM_1787307197_1 with a notable arrowhead shape. The cloud spectrum in Figure 5 shows enhanced brightness in all methane transmission windows down to 2 $\mu$m with the characteristic bumps in the 2.12 $\mu$m wing and 2.9 $\mu$m window, which are common properties of high Titan clouds (Brown et al. 2009, Griffith et al. 2000, Rodriguez et al. 2011, Turtle et al. 2018). The sea spectrum shows that the hydrocarbon seas have the strongest near-IR absorption and appear rather uniform across Titan's surface as seen in the inset image in Figure 5.

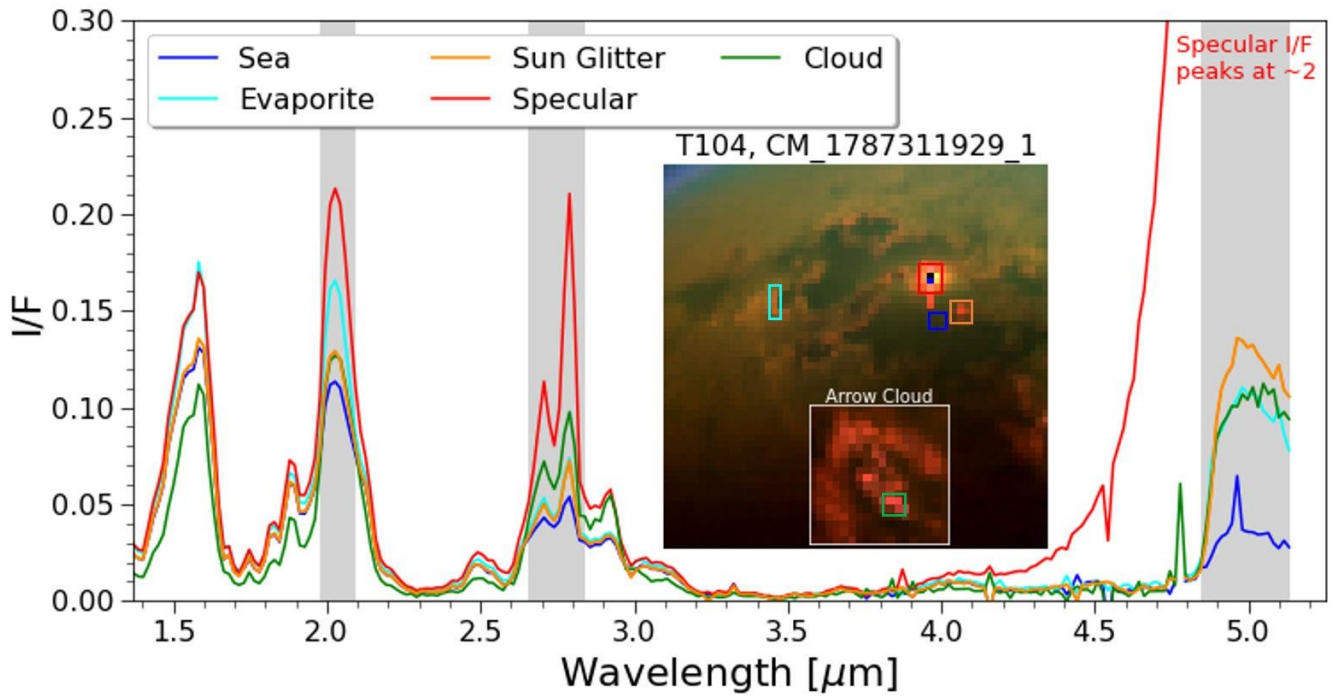

Figure 5. Near-IR spectra of surface and atmospheric features observed around Kraken Mare during the T104 flyby. Grey boxes indicate the transmission windows of Titan at 2.03, 2.75, and 5 $\mu$m. The inset plots show Kraken Mare from VIMS cube CM_1787311929_1 and the arrow cloud located north of Kraken Mare from cube CM_1787307197_1 with the same RGB color scheme as Figure 2. Spectral units are identified in the inset plot by boxes of the corresponding color. The spectra for evaporites and clouds show similar brightened I/F values for the 5 $\mu$m transmission window, while the hydrocarbon sea is the darkest surface feature. Evaporites are bright 5 $\mu$m deposits nestled along certain Kraken Mare coastlines. The brightest features are the non-saturated specular reflection and sun glitter observed in western Kraken Mare. The specular reflection pixel reaches a peak I/F of ~2.

The evaporite deposits are known to be bright at mainly 5 $\mu$m and have been documented along this coastal region of southeast Kraken Mare during many flybys (MacKenzie et al. 2014, MacKenzie & Barnes 2016). We note an increased 2 $\mu$m brightness of the evaporite spectrum in Figure 5. To our best knowledge, the surface phase function and composition of Titan evaporites remain unknown (MacKenzie & Barnes 2016). Nonetheless, we identify evaporites as a fixed coastal feature of Kraken Mare.

The 5 $\mu$m saturated pixels covering known areas of Kraken Mare are a clear sign of a specular reflection at a close spacecraft altitude (Soderblom et al. 2012). The brightest non-saturated pixel leaves no doubt of its specular nature as it goes off the chart in Figure 5. Also, there are reports of "bright ephemeral features" (BEFs) during north polar summer, possibly caused by wetted surfaces or fog (Dhingra et al. 2019, Dhingra et al. 2020). We show a T106 VIMS observation of a BEF in Figure 6 where there is potential sun glitter in Kraken Mare for direct comparison. We observe set of pink pixels near the specular reflection in Figure 6 but inside the sun glitter zone, which we attribute to the possible sun glitter. Conversely, we observe the large, documented wet-sidewalk feature (Dhingra et al. 2020) at a larger specular deviation angle of 14° in Figure 6. A comparison of their spectra in Figure 6 shows the BEF is dimmer than the sun glitter in the transmission windows. Thus, we are reasonably confident that the BEF is distinct from sun glitter from the T106 observation. The distinguishing features of the BEF include a large areal coverage over land and Ligeia Mare and a location outside of the sun glitter zone that is not consistent with rough sea surfaces.

This just leaves the sun glitter spectrum seen as the second brightest feature at 5 $\mu$m in Figure 5. Previous sun glitter observations from the T85 flyby (Barnes et al. 2014) likely only observed sun glitter at 5 $\mu$m due to strong forward-scattering at a high incidence angle of ~73° (Soderblom et al. 2012). Hence, the atmospheric scattering exceeds the surface reflectance for the sun glitter pixels, making the sun glitter indiscernible at wavelengths shorter than 5 $\mu$m. However, there is no physical reason why *Cassini* VIMS cannot view sun glitter at shorter wavelengths. Figure 5 confirms that the T104 sun glitter spectrum is brighter than the sea spectrum in the 2.03 and 2.75 $\mu$m windows at an incidence angle of ~55°.

## 4. FINE-SAMPLING SUN GLITTER OBSERVATIONS

We turn our attention to an analysis of fine-sampling (<10 km/pixel) observations of sun glitter in preferential areas of wave activity in the narrow straits (Freta) of Kraken Mare. We note that the relative location of the theoretical specular point becomes an issue of concern for fine-sampling observations. For moderate sampling (~20 km/pixel), the theoretical specular point is generally confined to a single pixel for the observation time of the cube. While closer in to Titan, however, we need to consider how the illumination geometry may change over the observation time of a fine-sampling VIMS cube as *Cassini* reaches altitudes of less than <15,000 km. The specular point trajectories and footprints of each fine-sampling cube in this section are shown in Figure 7 (Seignovert et al. 2020). A moving specular point can change the level of I/F enhancement of individual sun glitter pixels within a single cube. Accordingly, we address the specular point trajectory for each fine-sampling cube and the assumptions made on their "reliability" (i.e. ability to reasonably compare sun glitter features) in this section.

### 4.1. *Bayta Fretum*

*Cassini* VIMS collected hundreds of 32×1 "noodle" cubes during the T104 flyby that scanned over the entire Bayta Fretum region at a sampling of 7 km/pixel and were reconstructed with the steps outlined in Barnes et al. (2008). The T104 specular point moved over a distance of 35 km from the observation time of the first noodle pixel and the last sun glitter pixel (green ellipses in Figure 7). The travel distance is sufficiently small as to limit variations in illumination geometry of the sun glitter, so we can reasonably compare the Bayta sun glitter features. We note that the change of the spacecraft altitude is small and negligible for this observation.

Figure 8 displays the compiled noodle observations in wet-sidewalk color and the methane transmission windows. A cylindrical projection of the T104 noodle observation (top panel, Figure 8) shows a distinct coastline that correlates well with *Cassini* RADAR observations and three separate sun glitter features (orange arrows) in Bayta Fretum (Lopes et al. 2019). The two fainter sun glitter features directly north of Penglai Insula (top panel, Figure 8) only become visible at a higher sampling, unlike the largest linear sun glitter that represents the T104 Bayta sun glitter at moderate sampling (17 km/pixel) in Figure 4. Multiple sun glitter detections may imply that distinct wave fields originating from multiple sources are present in Bayta Fretum.

We comment on the largest fine-sampling sun glitter in the 4 transmission windows (2.03, 2.7, 2.79, and 5.0 $\mu$m) shown in the bottom panels of Figure 8. The structure of the sun glitter becomes non-uniform in specular point lies within the saturated zone. Bottom panel: VIMS near-infrared pixel spectra of the corresponding features. The pixel for each zone is outlined in the top panel. Grey regions indicate the methane transmission windows.

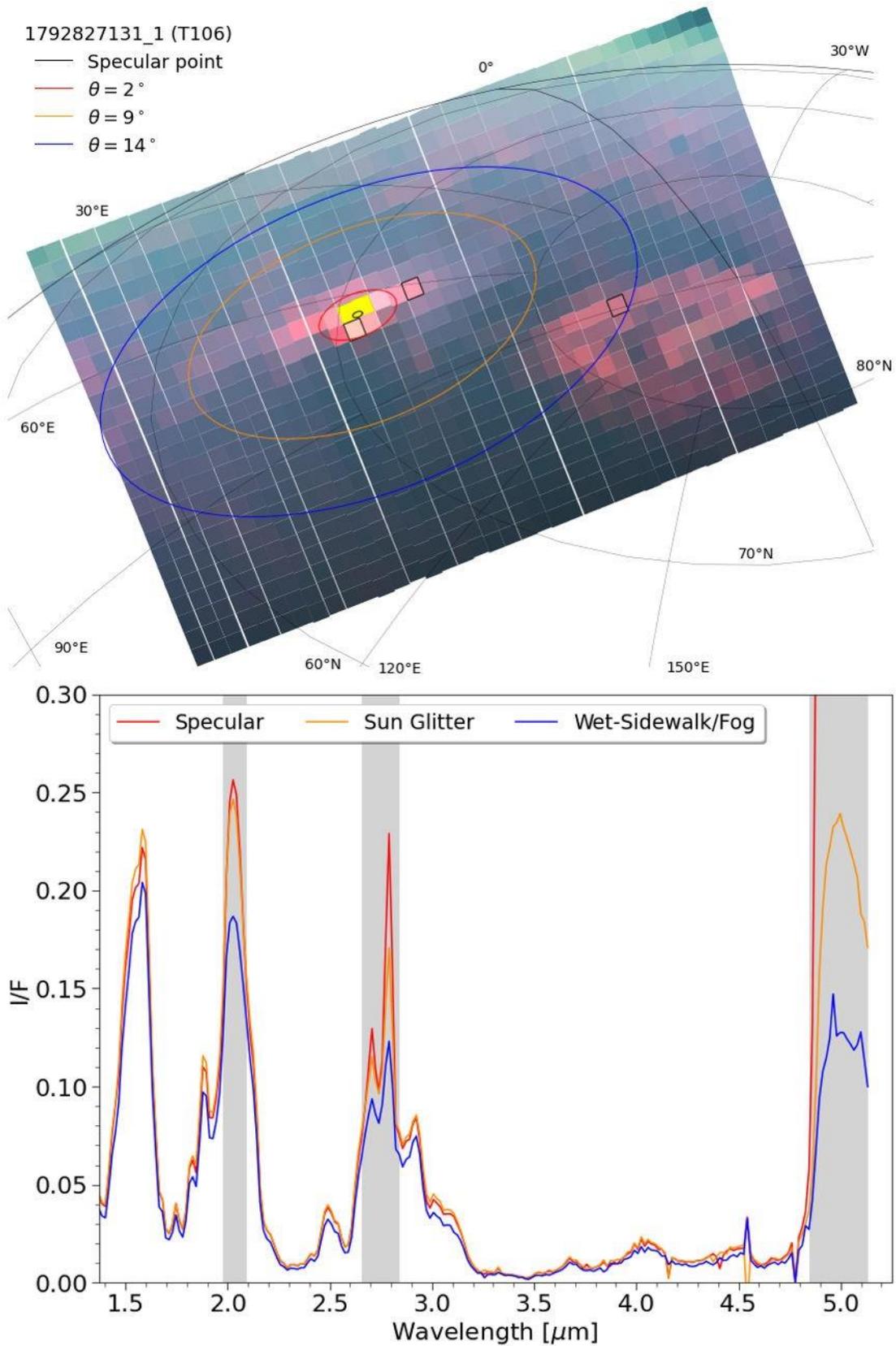

Figure 6. Top panel: The ortho-projected, saturated T106 VIMS observation with the wet-sidewalk RGB color scheme of Figure 2 contains both sun glitter and a wet-sidewalk feature (BEF). The ortho-projection did not use raster interpolation in order to preserve the features. The specular, sun glitter, and BEF zones are noted by red, orange, and blue ovals with their respective specular deviation angle. Yellow pixels indicate saturation. The black theoretical

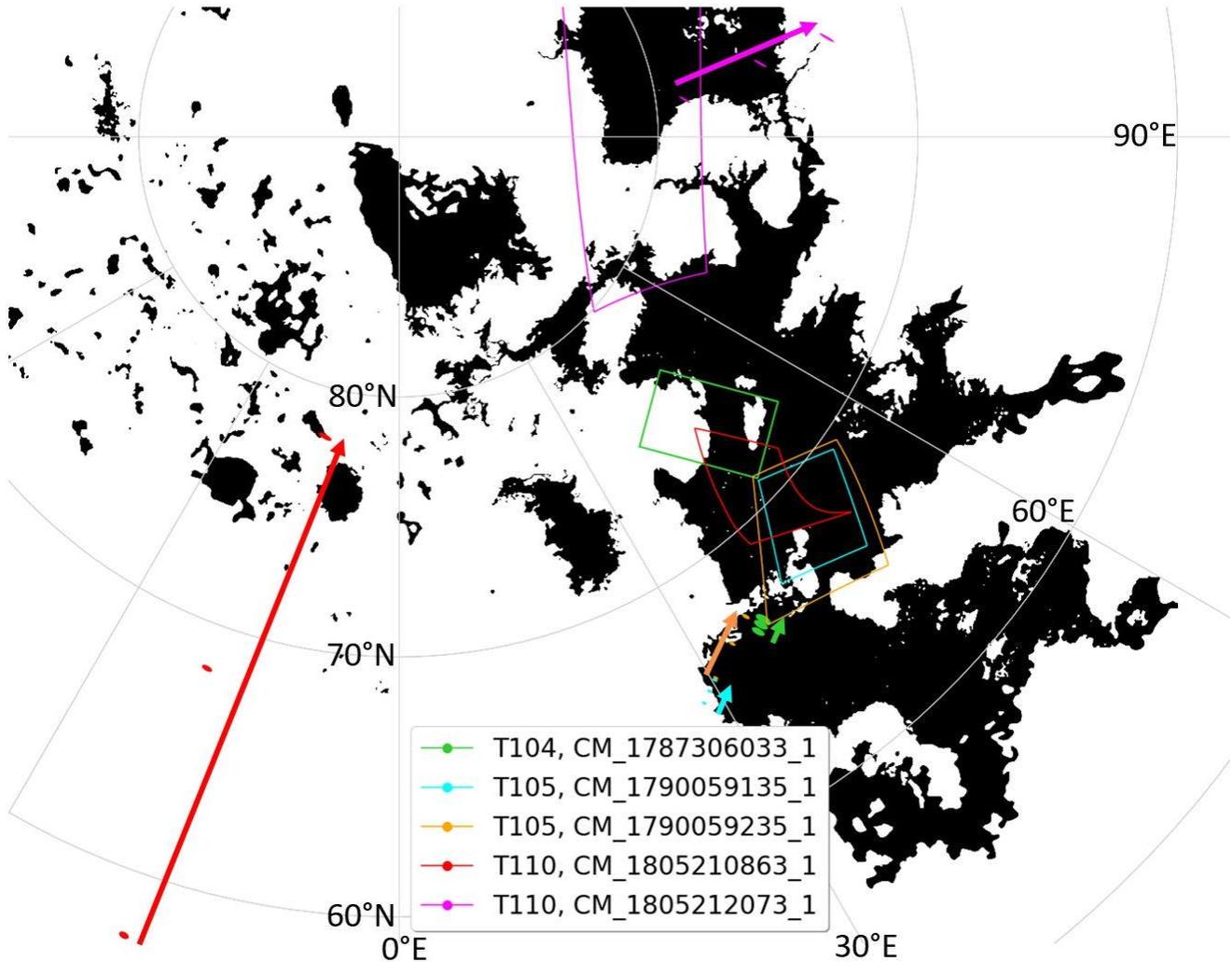

Figure 7. This figure showcases the fine-sampling VIMS cubes and their solid footprints from the T104, T105, and T110 flybys atop the binary map from Figure 3 in a polar projection. The colored ellipses show the relative dimensions and locations of the theoretical specular points over the observation time of the fine-sampling cubes. We plot the specular points at the start, middle, and end of the cube's observation time. The arrows indicate the direction and displacement of the specular point motion across the surface of Titan. Note that the specular point dimensions are 50 times larger for visibility, while red ellipses are 150 times larger.

the 5 $\mu$m window as seen by the red, dashed contours in Figure 8, which likely means that the wave field is variegated with some range of wave heights. The two brighter wave fields inside the largest sun glitter also appear at different distances between ~10-25 km from the coastline. We determine the approximate coastline at 1.58 $\mu$m where no sun glitter is present. We also observe the sun glitter with a similar morphology in the shorter atmospheric windows at 2.03, 2.7, and 2.79 $\mu$m. Note that the sun glitter I/F intensity varies with the atmospheric transparency of each window, where shorter wavelengths are less transparent. The sun glitter I/F is independent of the sea composition since the refractive indices of liquid methane and ethane show little change over VIMS wavelengths (Martonchik & Orton 1994), consistent with multi-wavelength sun glitter imagery of Earth's ocean (Kay et al. 2009).

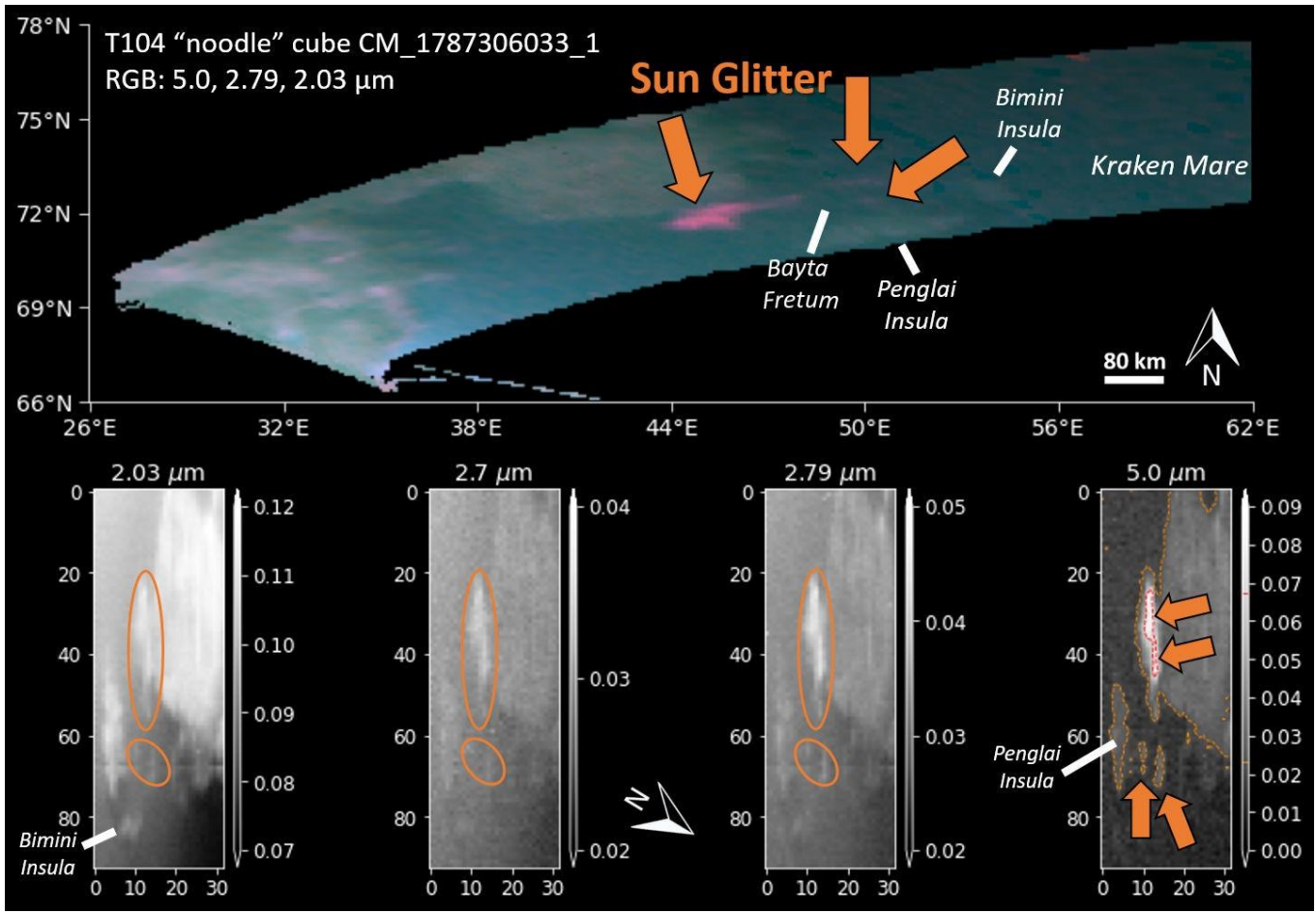

Figure 8. Top panel: A fine-sampling cylindrical map of the T104 "noodle" cube CM_1787306033_1 for northwest Kraken Mare with the same wet-sidewalk RGB color scheme as Figure 2. The observation reveals three separate sun glitter observations (orange arrows) in Bayta Fretum and along the northern coastline of Penglai Insula. Bottom panels: The unprojected "noodle" cube (green footprint in Figure 7) in the 4 transmission windows (left to right: 2.03, 2.7, 2.79, 5.0 μm) shows where the sun glitter pops out near the Bayta coastline. The 5 μm image includes red and orange dashed I/F contours of 0.023 and 0.057 respectively. The 5 μm image shows the far left orange contour as the Penglai Insula, while the smaller contours identify the three distinct sun glitter feature. The red contours identify notable wave fields in the largest sun glitter feature. At shorter wavelengths, orange ovals show the sun glitter features. The colorbar range highlights the relative transparencies of the transmission windows. Note that Bimini Insula is not visible in the 5 μm window due to its low signal-to-noise ratio (see Figure 3b of Sotin et al. 2012), but clearly appears at 2.03 μm.

During the T110 flyby, there was a fine-sampling (2 km/pixel) VIMS cube CM_1805210863_1 of Bayta Fretum that overlapped the largest T104 sun glitter in Figure 9. We note that the brightened surface on Kraken Mare is not caused by a specular reflection but rather skylight being reflected off the sea surface due to emission angles exceeding 60° (Vixie et al. 2015; refer to Figure 5 of Barnes et al. (2014) for a visual example). The right image of Figure 9 seemingly shows no evidence of surface roughness amongst the brightened sea surface in the T110 observation. On a brightened sea surface, we expect surface roughness to manifest as dark patches since they limit the number of surface facets able to reflect skylight. However, we are unable to fully rule out the possibility of surface roughness since VIMS does not have the sensitivity to distinguish darker regions on a sea surface with very low reflectance (~1%) (Sotin et al. 2012). In addition, the T110 specular point is located several hundred kilometers from Bayta Fretum (Figure 7), preventing the detection of any possible sun glitter.

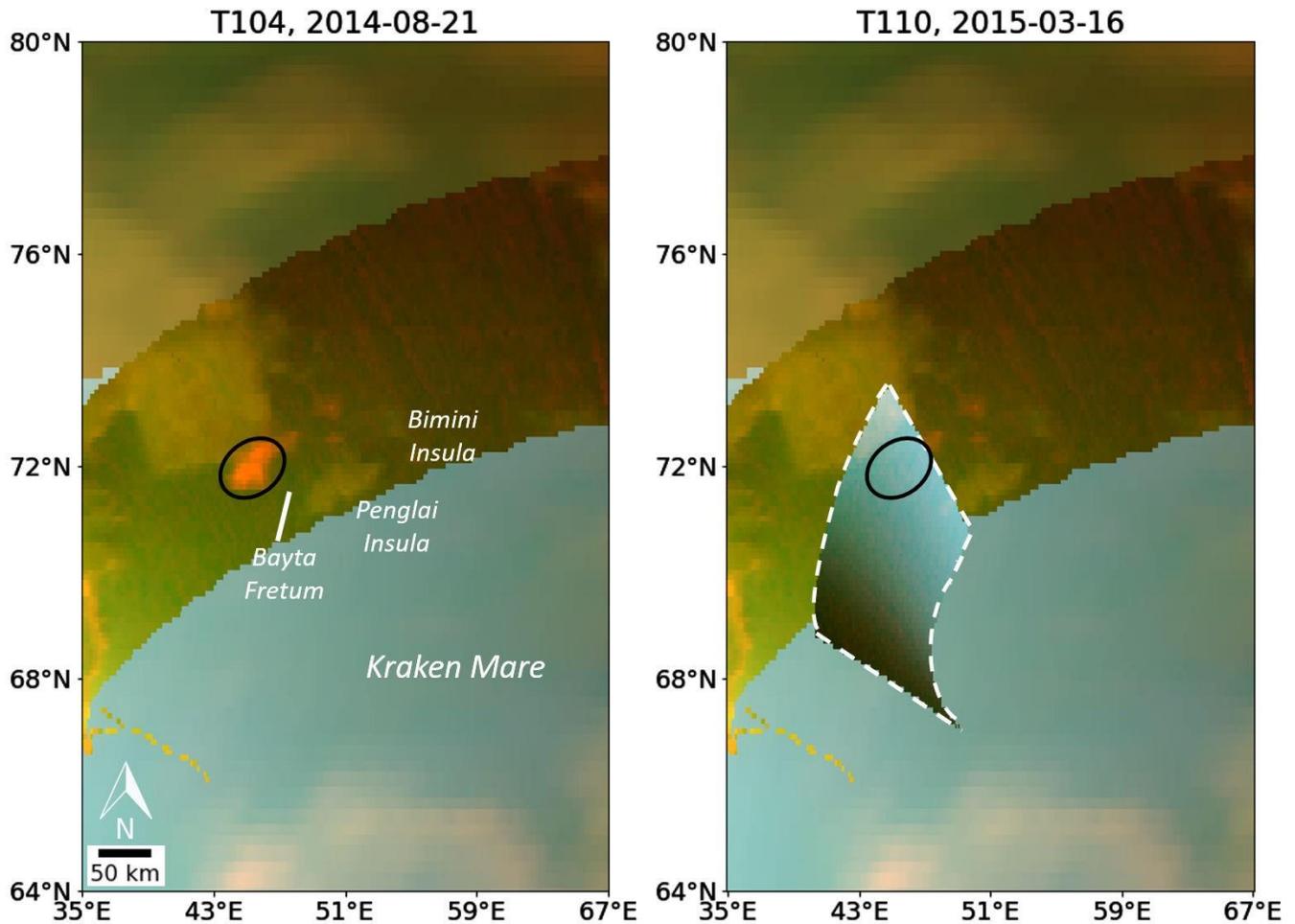

Figure 9. The left image shows the same fine-sampling "noodle" T104 cube Figure 8 overlaid on a cylindrical basemap of northwest Kraken Mare with the same contrast-enhanced wet-sidewalk color scheme as Figure 2. The right image shows the VIMS cube CM_1805210863_1 (white, dashed lines) from the T110 flyby that overlies the T104 Bayta sun glitter region from the left image. The diffuse reflection on the sea is due to emission angles exceeding 60° (Vixie et al. 2015). This observation shows no sun glitter during the T110 flyby in Bayta Fretum; however, the observation geometry for this cube limits our ability to rule out surface roughness.

### 4.2. *Seldon Fretum*

The 40×17 km Seldon Fretum and nearby archipelago have been suggested to play a significant role in the tidal interactions and liquid exchanges between the north and south Kraken Mare basins (Lorenz et al. 2014, Tokano 2010, Tokano et al. 2014, Vincent et al. 2018). Fine-sampling sun glitter imagery can therefore shed light on the sea surface dynamics of the Seldon Fretum region. As such, we present the first observational evidence of sun glitter in 2 separate fine-sampling (2-3 km/pixel) cubes, CM_1790059135_1 and CM_1790059235_1, of the immediate region of Seldon Fretum during the T105 flyby. The T105 specular point travels 115 km northeastward over the observation time of VIMS cube CM_1790059235_1 (orange

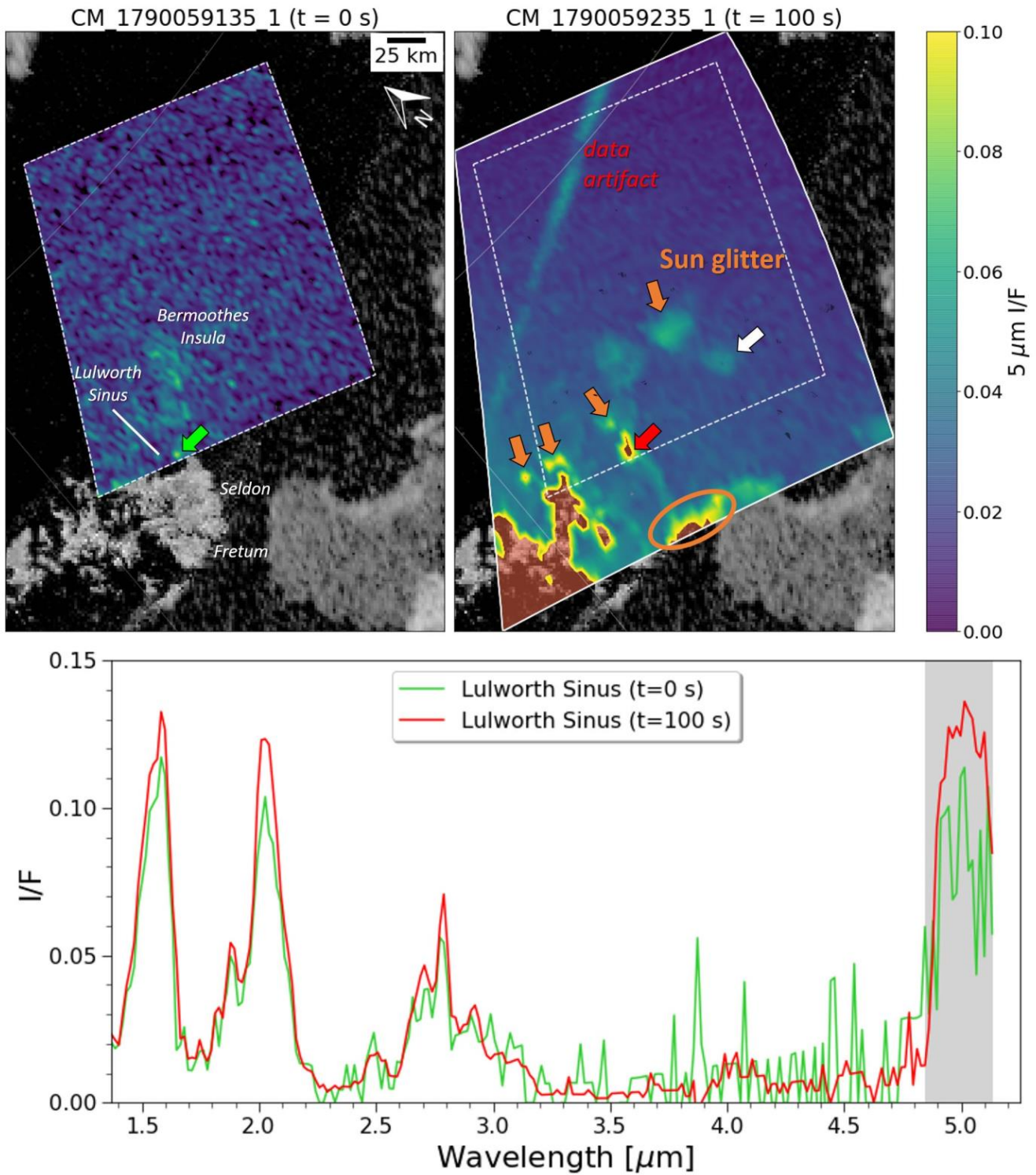

Figure 10. Two fine-sampling (∼2 km) VIMS observations of sun glitter near Seldon Fretum from the T105 flyby in a polar stereographic projection overlying a *Cassini* HiSAR RADAR basemap (Lopes et al. 2019). The white outlines indicate the spatial extent of each cube. The red color indicates a 5 μm I/F greater than 0.1 that corresponds to the specular "zone" with a maximum I/F of ∼0.76. These observations were taken 100 seconds apart. The second observation (t=100 s) has an 80 ms exposure time. We observe numerous instances of sun glitter (orange arrows) with the brightest ones occurring in Lulworth Sinus and Seldon Fretum (orange oval). Weaker sun glitter was detected near the north tip of Bermoothes Insula. The white arrow points to offshore sun glitter. The first cube was taken at a 20 ms exposure time, resulting in a poor signal-to-ratio for the 5 μm window. Thus, there was only one repeat

detection of sun glitter between the two observations in Lulworth Sinus, based on its overall I/F enhancement at 5 $\mu$m (grey region) shown in the bottom spectral plot.

ellipses in Figure 7). So, we do not include pixels with an I/F greater than 0.1 from the specular zone, denoted by the red region in Figure 10, from this cube in our analysis.

Overall, we identify six isolated instances of sun glitter in CM_1790059235_1, denoted by arrows or ovals in Figure 10. We overlay a *Cassini* RADAR basemap (Lopes et al. 2019) to put these instances of sun glitter into geographic context. We note that the red region likely contains sun glitter in the narrow channels between the islands.

First, we observe a notable extent of sun glitter that nearly overlays the northern terminus of Seldon Fretum (orange oval in Figure 10). The striking part of the Seldon sun glitter is how it terminates at the northern edge of the strait, showing how the sea surface roughness is confined to Seldon Fretum.

Second, the entrance of a previously unnamed cove has notable sun glitter that exceeds a 5 $\mu$m I/F of 0.1 (red arrow in Figure 10). This cove, Lulworth Sinus, has a narrow 10 km mouth from RADAR data (Lorenz et al. 2014), so the Lulworth sun glitter indicates notable sea surface roughness at the mouth. In addition, the Lulworth sun glitter was the only one likely observed in a second fine-sampling VIMS cube, CM_1790059135_1, taken only 100 seconds earlier (green arrow in Figure 10). The specular point distance for the Lulworth sun glitter pixel in cube CM_1790059135_1 is ∼300 km, so we can reasonably compare the Lulworth sun glitter pixels between the two T105 cubes. Unfortunately, the 20 ms exposure time of cube CM_1790059135_1 created a rather noisy observation at 5 $\mu$m (top left plot of Figure 10). Nonetheless, we observe an overall I/F increase in the 5 $\mu$m window for both Lulworth sun glitter pixels in their spectra in Figure 10. Overall, we cannot be certain of a time-resolved observation of sea surface roughness at the mouth of Lulworth Sinus without ruling out the different viewing geometry between the two cubes.

Third, we note four isolated instances of sun glitter near tiny RADAR-bright islands off the coasts of Hufaidh Insulae and Bermoothes Insula (orange arrows in Figure 10). The locations of this sun glitter near the coastlines of tiny islands may suggest a build-up of sea surface waves (i.e. a wave field). If this interpretation of the sun glitter is true, then the wave fields may originate from multiple sources, given their relative locations on different sides of their respective island. The rightmost sun glitter (white arrow in Figure 10) is not located near any visible coastline, suggesting local sea surface roughness may be caused by unseen factors, such as a submarine landmass (e.g. a seamount) disrupting a surface current.

Finally, we address the presence of a ∼100 km linear feature that appears in cube CM_1790059235_1 (denoted as "data artifact" in Figure 10). The specular point trajectory for the observation time of this cube is directly behind the cube's orange footprint in Figure 7. We also do not see this alignment or similar artifacts in any other fine-sampling VIMS observation. Thus, we attribute the feature to a sunbeam (Lynch & Livingston 2001) radiating from the specular point. This artifact is similar to previous specular reflection anomalies (Barnes et al. 2013) caused by an instrumental artifact due to internal reflections in VIMS.

### 4.3. *Tunu Sinus*

The ∼20 km wide strait Tunu Sinus has also been a place of interest in regards to the fast tidal flows predicted for this portion of the north Kraken Mare basin (Tokano 2010, Vincent et al. 2018). We observe evidence of sun glitter present in Tunu Sinus in a fine-sampling (3 km/pixel) cube, CM_1805212073_1, during the T110 flyby, shown in Figure 11. The T110 specular point only moves 25 km southward over the observation time of the Tunu Sinus region (purple ellipses in Figure 7), allowing us to compare the individual sun glitter features.

Four or five unique instances of 5 $\mu$m sun glitter appear confined to the narrow straits or concave coastlines of Tunu Sinus. Specifically, we note the brightest sun glitter coincides with the narrowest (∼15 km) portions of Tunu Sinus, which may correspond to the most turbulent wave fields in this observation.

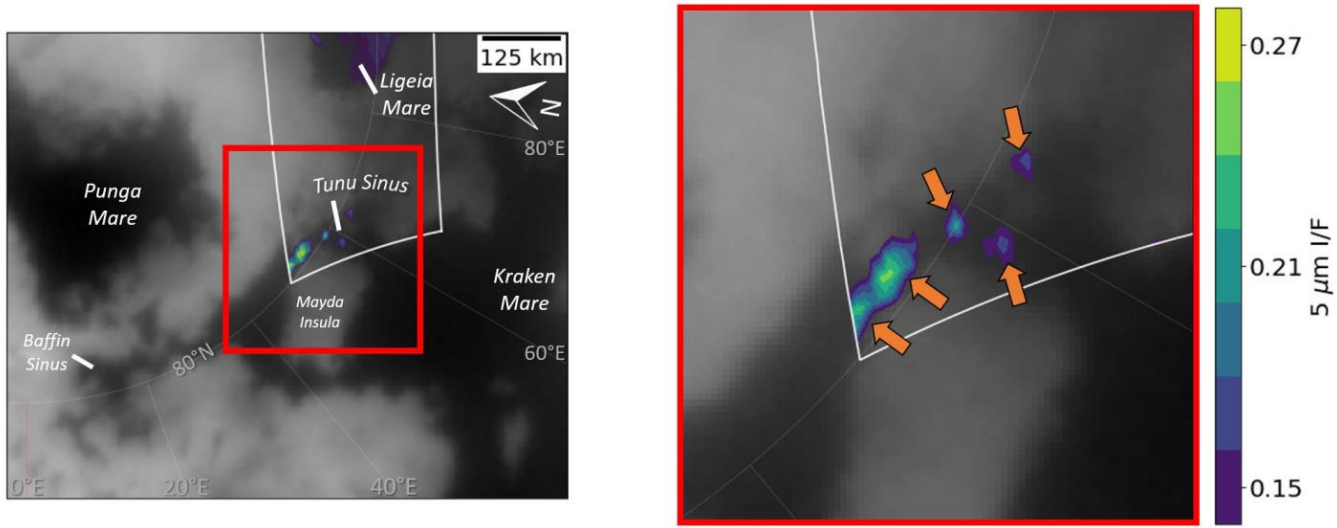

Figure 11. The left image show the 2.8 km² *Cassini* ISS basemap (Karkoschka et al. 2017) in a polar projection with labeled features of the maria. The right plot shows an inset image (red box) of the fine-sampling (∼3 km) T110 VIMS observation CM_1805212073_1 with isolated sun glitter in Tunu Sinus at 5 *μ*m, indicated by orange arrows. The white outline indicates the spatial extent of the VIMS image cube. The I/F range isolates the sun glitter in Tunu Sinus since the 5 *μ*m I/F is ∼0.1 for the dark sea pixels.

## 5. DISCUSSION

We postulate as to the possible origins of the wave fields associated with the sun glitter observations.

### 5.1. *Origin of T104 Bayta Fretum sun glitter*

Large swaths of sun glitter are common features for various coastlines of Earth from airplanes and finesampling remote sensing (Kay et al. 2009). Deep-sea surface waves approaching a coastline experience wave-shoaling or increases in wave height from turbulent actions with the seafloor (Craik 2004). The wave base is the maximum depth at which a propagating surface wave can turbulently interact with the seafloor. On Earth, the wave base is considered one-half of the wavelength of a surface wave. Thus, the variegated sun glitter may indicate a shallow bathymetry for Bayta Fretum. However, this explanation does not pin down the origin of the wave fields observed in Bayta Fretum. Nonetheless, we postulate two possible origins for deep-sea waves reaching Bayta Fretum with inference from applicable physical models of Titan's sea processes: 1) tidal currents and 2) wind swells.

First, tidal models of Titan have attempted to estimate the general trends of tidal surface currents in the northern maria over a Titan day (Tokano et al. 2014, Vincent et al. 2018). The most recent model predicts that the wider margins of Bayta Fretum might only allow for a maximum southward tidal current of 1 cm/s at perikron, Titan's closest approach to Saturn (Vincent et al. 2018). This observation generally holds for several assumed bathymetric profiles (Vincent et al. 2018), considering the bathymetry of Kraken Mare is mostly unknown (Hayes et al. 2018). Tidal currents of 1 cm/s do not reach the 11 cm/s threshold predicted for capillary-gravity wave generation (Hayes et al. 2013). Furthermore, we note that the T104 and T110 Bayta observations from Figure 9 were recorded at the same true anomaly, indicating diurnal tides may not be responsible for the sun glitter. As a result, a coastal tidal current likely does not explain the Bayta wave fields without an additional unknown factor, such as a local area of shallow bathymetry.

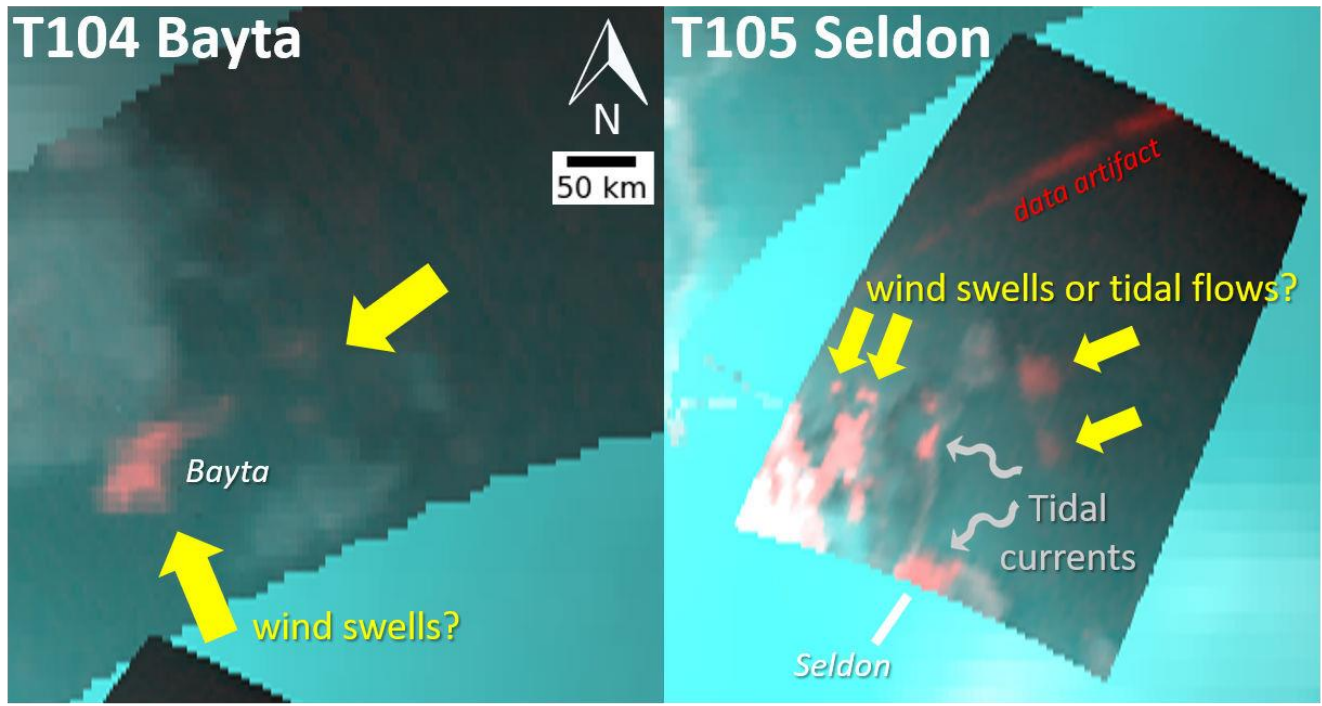

Figure 12. Summary of the possible origins of the fine sampling sun glitter and associated wave fields in the T104 noodle cube and T105 cube CM_1790059235 1. We use the same wet-sidewalk RGB color scheme of Figure 2. We note the linear sunbeam as the "data artifact" in the T105 cube. The arrows are not an indicator of flow direction, which remains unknown. The largest Bayta wave fields likely originate from winds given the weak tidal currents predicted in Bayta Fretum. For Seldon Fretum and Lulworth Sinus, the wave fields seem to originate from the constriction of tidal currents in their mouths. The other weaker sun glitter may originate from the winds or tides with enhancement from local areas of shallow bathymetry near the coastlines (i.e. wave-shoaling).

Secondly, wind-induced wave activity has been long-speculated on Titan's maria (Ghafoor et al. 2000, Ori et al. 1998) considering the ubiquitous occurrence of surface waves on Earth, which ranges from capillary ripples to 30 m high waves in the open ocean (Toffoli et al. 2017). Initial laboratory experiments and several wind-wave models of Titan have found that the maximum predicted 1 m/s winds can generate capillary wind waves with meter-scale wavelengths and wave heights of ∼20 cm (Ghafoor et al. 2000, Hayes et al. 2013, Lorenz et al. 2005, Lorenz & Hayes 2012, Lorenz et al. 2012a). Furthermore, coupled sea-GCM_models indicate relatively ethane-rich seas, like Kraken Mare, might enhance moist convection and subsequent wind activity directly over the seas (i.e. a sea breeze) (Tokano et al. 2009). Our moderate-sampling T104 and T105 sun glitter observations show evidence of consistently rough sea surfaces in Bayta Fretum, suggesting the possible wind source has a regional origin within the north Kraken basin.

The T104 observations of the arrow cloud located north of Ligeia Mare (Figure 5) and a northern Ligeia "magic island" interpreted as waves (Hofgartner et al. 2016) may indicate the arrow cloud is associated with deep convective processes and stormy weather over Ligeia Mare. However, the Bayta waves are ∼700 km away from the arrow cloud. Gust fronts formed by storm downdrafts may generate surface winds of 10 m/s (Charnay et al. 2015), but likely only have a radius of influence of a few hundred kilometers based on numerical storm simulations in Titan's lower troposphere (Rafkin & Barth 2015) and terrestrial observations (Lothon et al. 2011). Thus, we cannot ascertain that the arrow cloud system could have wind fields strong enough to generate wind swells out at Bayta Fretum. In addition, Bimini and Penglai Insulae may act as large breakwaters that protect Bayta Fretum from large wind swells.

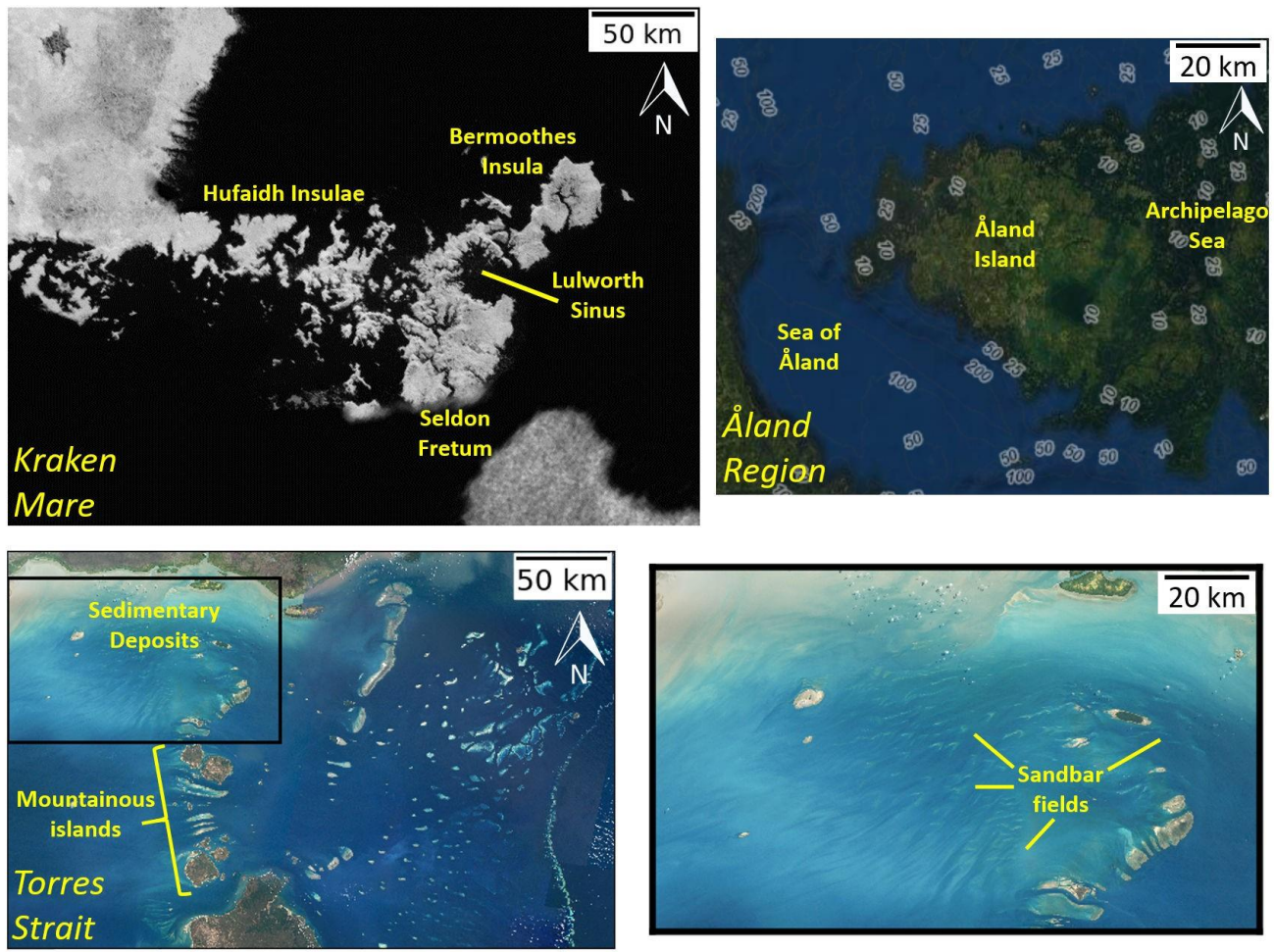

Figure 13. Two possible scenarios for the marine environment of Bermoothes and Hufaidh Insulae with terrestrial analogues. Top left panel: A cleaned *Cassini* RADAR map of Hufaidh and Bermoothes Insulae in a cylindrical projection (Credit: NASA/JPL-Caltech/Space Science Institute/Ian Regan). Top right panel: An ESRI World Imagery basemap of the Aland Region situated between Sweden and Finland with local NOAA bathymetry depths in meters. Bottom left and right panels: A compiled NASA MODIS mosaic of the Torres Strait in a cylindrical projection (Lawrey 2013) and a subwindow of the north region (black ox) respectively. This Kraken Mare archipelago displays a similar island density and coastal morphology to the Aland Region and Torres Strait. Both terrestrial analogues host shallow bathymetry. This may imply a similar bathymetry for Bermoothes and Hufaidh Insulae if there was a recent geological period of sea-level rise or tectonic uplift. If a large sediment supply is available in Bermoothes and Hufaidh Insulae, then tidal currents may form submarine features similar to the intricate sandbars seen in the northern Torres strait (bottom right panel).

Overall, the differentiation between surface waves induced by tides or wind is quite difficult with limited fine-sampling observations and observation geometries. Sea-wind circulation models show that the coastal regions of Kraken Mare may experience similar sea level changes from the tides and summer winds (Tokano & Lorenz 2015). Specifically, the maximum wind setup (i.e. increase in sea level near the coast due to wind-wave action) is nearly equivalent to the tidal range of ~8-10 cm on the Bayta shoreline. Thus, the combination of our sun glitter observations and model predictions suggest a complex dynamic for wave activity initiated by the winds or tides that depends on unknown coastline characteristics, such as bathymetry (Hayes et al. 2018).

### 5.2. *Origin of T105 Seldon Fretum sun glitter*

Seldon Fretum is a narrow strait connecting the two large basins of Kraken Mare expected to host choppy seas (Lorenz et al. 2014).

#### 5.2.1. *Tidal Currents: Seldon Fretum and Lulworth Sinus*

The most noteworthy sun glitter detection in the fine-sampling T105 cube CM_1790059235_1 were in Seldon Fretum and the entrance of Lulworth Sinus. Seldon Fretum is the location where fast northward tidal currents are predicted in tidal models at ~0.2 Titan days after apokron (Tokano et al. 2014, Vincent et al. 2018). The tidal models assumed shallow bathymetric profiles (<50 m) for Seldon Fretum and Lulworth Sinus, based on the idea that depth scales with distance from the shoreline (Lorenz et al. 2014). Thus, we find the sun glitter in these regions consistent with tidal currents constricted at the strait entrances that manifest as turbulent sea surfaces. We quantify the possible sea surface roughness in these narrow straits. The Reynolds number helps describe the type of flow for a liquid in a channel, given by:

$$Re = \frac{4uA}{\nu P} \tag{1}$$

where $u$ is the flow speed, $A$ is the channel cross-sectional area, $P$ is the wetted perimeter, and $\nu$ is the kinematic viscosity. If we assume a rectangular channel width and depth of 20 km and 50 m respectively with liquid methane flowing at decimeter-scale tidal current velocities predicted in models (Vincent et al. 2018), Seldon tidal currents easily reach the turbulent flow regime ($Re > 2300$). Also, energy balance ($gh = 0.5u^2$) shows that 0.3 m/s tidal current velocities ($u$) in Seldon Fretum can cause a sea surface roughness ($h$) of ~3 cm, reaching the capillary-gravity wave threshold (Hayes et al. 2013). The surface liquid exchanges between the Kraken basins would be driven by tidal flows with only minor influences from surface winds in Seldon Fretum (Lorenz et al. 2014, Ori et al. 1998). We anticipate a spectrum of centimeter-scale wave heights in Seldon Fretum, but constraining wave heights remains beyond the scope of this work without bathymetric profiles.

The occurrence of internal waves, the surface manifestations of turbulence at the boundary of two stratified layers in the sea, has been suggested for Seldon Fretum due to the possibility of stratified seas (Lorenz et al. 2014). Recent theoretical and experimental evidence shows one possible scenario for a two-liquid layer stratification with a methane-rich top layer and an ethane-rich bottom layer in Kraken Mare (Cordier et al. 2017, Hanley et al. 2018). If this binary liquid were present in Seldon Fretum, then tidal currents in the top methane-rich layer could cause roughness on the bottom ethane-rich layer and subsequently generate internal waves that propagate to the surface at some distance from Seldon Fretum. However, if Seldon Fretum has shallow bathymetry, we note that any stratification may be broken due to significant mixing from the bottom friction of the seafloor. On Earth, there is a lack of internal waves in certain straits, such as the English channel (Jackson 2007). Alternatively, a shallow Seldon Fretum may only allow for the exchange of the surface methane layer, also preventing internal wave formation.

#### 5.2.2. *Surface waves: Bermoothes and Hufaidh Insulae*

The coarse-sampling T105 cube in Figure 4 shows a large swath of the surface of north Kraken Mare being roughened, which we interpret as windy surface conditions over Kraken Mare on this Titan day. The 4 separate detections of sun glitter along the northern shores of Bermoothes and Hufaidh Insulae (orange arrows in Figure 10) have two possible origins: winds or tidal currents. In practice, distinguishing between the two is quite hard from a single sun glitter observation. We are likely only observing a small part of the actual wave angle spectrum that is limited by our viewing geometry bias. We note that one sun glitter feature is located ~30 km offshore of Bermoothes Insula (white arrow in Figure 10), which may indicate wind-roughened seas in Kraken Mare. The red regions over Hufaidh Insulae in Figure 10 are likely indicative of surface waves generated from constricted tidal currents. Overall, we cannot determine a single origin for these Insula wave fields but rather likely originate from a combination of the tides and summer winds.

### 5.2.3. *Coastal Morphology of Hufaidh and Bermoothes Insulae*

We can interpret the coastal morphology and geology of Hufaidh and Bermoothes Insulae using the observations of various wave fields in the Seldon Fretum region. We identify two potential terrestrial analogues to this region: the Aland Region in the Baltic Sea (Lorenz et al. 2014) and the Torres Strait between Australia and Papua New Guinea. These terrestrial analogues are both island-dense straits with shallow bathymetry (<25 m) but different sediment supplies available in their submarine environment.

First, the ~160 km wide Torres Strait that divides the north tip of Queensland, Australia and Papua New Guinea shows an array of islands, more densely-populated in the south, and sediment-rich region to the north in the bottom panels of Figure 13. The Torres Strait demonstrates some of the most complex tidal circulation trends on Earth from the interactions of two distinct ocean basins with different general circulation patterns and mean sea levels (Lemckert et al. 2009, Wolanski et al. 1988). In addition, the bathymetric profile of the Torres Strait ranges from 7-15 meters (Harris 1988). The complex tidal currents and shallow bathymetry result in highly localized variations in wave activity in the Torres Strait (Hemer et al. 2004), which may be an analogous scenario for the straits of Hufaidh and Bermoothes Insulae. The tidal action becomes apparent from the wave-like bedforms, such as sand waves (Daniell et al. 2015, Pilkey et al. 2011) in the bottom right image of Figure 13, analogous to sand dunes sculpted by the winds. We suggest a version of "icy" sandbars as an intriguing possibility for the submarine environment of Hufaidh Insulae, previously postulated by Lorenz et al. (2003) and Lorenz (2013). This scenario would only hold if there was a large sediment runoff from nearby rivers dumping into Kraken Mare (Burr et al. 2006, Ori et al. 1998).

Second, the Aland region has been suggested as a terrestrial analogue to Hufaidh Insulae, Bermoothes Insula, and Seldon Fretum (Lorenz et al. 2014). The top right panel of Figure 13 shows the Aland region and equivalent Titan features, where the Sea of Aland is Seldon Fretum, the Aland Island is Bermoothes Insula, and the Archipelago Sea is Hufaidh Insulae. We observe complex daily surface circulation patterns in the Archipelago Sea (Erkkila & Kalliola 2004 ̈ ) with shallow depths (Figure 13), which may share the circulation patterns of the Hufaidh Insulae straits. The Sea of Aland often hosts choppy seas (Kahma et al. 2003), which may resemble the wave conditions of Seldon Fretum. Meanwhile, the Aland and Archipelago Sea islands host a mix of cliffed and ria/flooded coastlines in a similar manner to Bermoothes and Hufaidh Insulae respectively. However, the key difference from the Torres Strait is the lack of sedimentary deposits and subsequently no submarine bedforms.

For the actual islands of Hufaidh and Bermoothes Insulae, geomorphological analysis shows the region is likely composed of the hard primordial crust that predates the hydrocarbon seas (Birch et al. 2017). The rugged, sinuous coastlines and internal river valleys may resemble the submergent coastlines of Earth with rias or flooded coastal inlets (Bird 2011). The rugged coastlines suggest a slow erosion rate of the Insulae, similar to the erosional history of Mayda Insula (Lucas et al. 2014). The submergent coastlines of Hufaidh and Bermoothes Insulae might also hint at a similar geological history to the two terrestrial analogues. For the Torres Strait, the local peaks of the southern Torres Strait islands were likely isolated from rising sea levels after the Last Glacial Period (Allen et al. 1977). For the Aland Region, the hard granite islands likely rose out of the sea due to post-glacial isostasy from the last ice age (Edelman & Jaanus 1980). Thus, the local peaks in the Hufaidh and Bermoothes Insulae may originate from either: a period of sea-level rise or tectonic uplift of an erosion-resistant bedrock.

One final geological feature is the semi-circular Lulworth Sinus. The observation of bright sun glitter at the entrance to Lulworth Sinus in two separate T105 cubes indicates tidal currents and considerable wave activity in Lulworth Sinus. The interior coastlines appear to host several rias in Figure 13, suggesting that Lulworth Sinus was likely infilled during a period of sea-level rise or isostatic subsidence. Thus, wave action likely has not played a role in the formation or semi-circular shape of Lulworth Sinus.

### 5.3. *Origin of T110 Tunu Sinus sun glitter*

Stereo topography of the nearby Mayda Insula from *Cassini* RADAR suggests significant fluvial erosion of the terrain (Lucas et al. 2014). Tidal models of Titan predict decimeter-scale tidal flow velocities in the Tunu Sinus region (Tokano et al. 2014, Vincent et al. 2018). We suggest that the sun glitter likely originates from the constriction of tidal currents flowing into the narrow straits and concave coastlines of Tunu Sinus. We are unable to rule out the influence of winds without an observation of local weather conditions during the T110 flyby due to the high phase angle. Fluvial erosion may be an ongoing geological process driven by tidal flows in the coastal areas of Tunu Sinus (Ori et al. 1998). The observation of active tidal currents in Tunu Sinus, combined with evidence for the Maria sharing an equipotential surface (Hayes et al. 2017), may also indicate active hydrological connections between Kraken and Punga Maria via surface channels currently unresolved by RADAR.

## 6. CONCLUSION

The T104 and T105 VIMS observations show evidence of isolated sun glitter across the surface of Kraken Mare at moderate sampling (17-21 km/pixel) that indicate surface waves. Fine-sampling observations (<10 km/pixel) during the same flybys find evidence of variegated wave fields and confirm the moderate-sampling sun glitter over three narrow straits: Bayta Fretum, Seldon Fretum, and Tunu Sinus. Geographic projection of the VIMS observations and spectral analysis indicate that the sun glitter is confined to the sea surfaces and distinct from other known surface and atmospheric features, namely clouds/fog, potential rain-wetted surfaces, and evaporites.

Fine-sampling sun glitter imagery from the T104 flyby reveals an extended set of wave fields that are parallel with the Bayta Fretum shorelines. While an arrow cloud and Ligeia surface waves observed during T104 provide evidence for convective weather, we cannot determine if storm-induced winds are solely responsible for the Bayta waves. Overall, our observations suggest complex influences for wave activity in Bayta Fretum, possibly originating from the winds. The tidal currents in Bayta Fretum are unlikely to reach the capillary-gravity wave generation threshold. Overall, we cannot be certain of tidal or wind wave action in Bayta Fretum without knowledge of its bathymetry.

We observe several instances of sun glitter in the Seldon Fretum region in two T105 fine-sampling VIMS observations. In particular, Seldon Fretum and the entrance of Lulworth Sinus host the brightest sun glitter, which we attribute to the constriction of northward tidal currents predicted by tidal models (Tokano et al. 2014, Vincent et al. 2018) for a true anomaly of 246° (Table 1). The T105 Hufaidh and Bermoothes Insulae sun glitter suggest a complex interaction of wind and tidal activity. However, one instance of sun glitter ∼30 km offshore from Bermoothes Insula may indicate a detection of wind seas over Kraken Mare during the T105 flyby or an anomalous feature underneath the sea surface that interrupts surface currents. Internal waves are an unlikely prospect for a Seldon Fretum with shallow bathymetry.

We infer about the coastal morphology of Hufaidh and Bermoothes Insulae from the T105 Seldon wave observations. The Torres Strait and the Aland region may serve as analogous features to Hufaidh and Bermoothes Insulae, considering they are both shallow straits with many islands that experience complex surface circulation and wave patterns. However, there is a notable difference in their marine environments as the Torres Strait hosts an abundant sediment supply from Papua New Guinea rivers with unique seafloor bedforms (e.g. sandbars) formed by tidal action. Hufaidh and Bermoothes Insulae could have submarine sedimentary features, such as "icy" sandbars. Alternatively, Hufaidh and Bermoothes Insulae may simply host a mix of cliffed and ria coastlines with no major sedimentary deposits, just like the Aland region. Also, the islands of Hufaidh and Bermoothes Insulae are likely isolated peaks of the primordial Titan crust due to either the drowning of local peaks from a period of sea level rise or local tectonic uplift. These scenarios match the geological origins of the Torres Strait and the Aland region respectively. Finally, the shorelines of Bermoothes Insula and Lulworth Sinus may show evidence of a submergent coastline with flooded river valleys or rias and imply a recent episode of sea-level rise for Kraken Mare. Tunu Sinus tidal flows and the common equipotential surface of Titan's Maria may lean evidence toward a surface hydrological connection via unresolved rivers between Kraken and Punga Maria.

Future Titan missions could constrain the optical properties of Kraken Mare in order to pin down the true surface reflectance of sun glitter and its source. In particular, a Titan orbiter needs to take careful preparation for sun glitter imagery to eliminate artifacts. Also, a Titan orbiter with a wide-field camera could remove ambiguities of the viewing geometry in future observations by observing the same area of sun glitter over an orbit and many incidence and emergence angles. Acquiring constraints on the environmental conditions of the maria would prove beneficial for determining the mission operations and physical limitations of a Titan submarine (Hartwig et al. 2016). The recently-announced *Dragonfly* mission could use aerial images of potential local pockets of surface liquids from rainstorms to deduce surface winds, as has been demonstrated by airplanes and drones on Earth (Yurovskaya et al. 2018a, Yurovskaya et al. 2018b).

MFH, JWB, and JMS acknowledge funding from NASA *Cassini* Data Analysis and Participating Scientists (CDAPS) grant NNX15AI77G. We thank Rose Palermo and Andrew Ashton for their useful discussion on coastal geomorphology. We acknowledge useful manuscript recommendations from Vincent Chevrier and an anonymous reviewer. The authors acknowledge support from the NASA/ESA *Cassini* Project in providing data from the VIMS instrument.